\documentclass[slac_one]{revtex4}
\usepackage{graphicx}
\usepackage{fancyhdr}
\newcommand\new{\newcommand}         
\new{\Bs}  {{\ifmmode \mathrm{B}^0_s       \else $\rm B^0_s$ \fi}}
\new{\Bsbar} {{\ifmmode \overline{\mathrm{B}}^0_s \else $\overline{\mathrm{B}}^0_s$ \fi}}
\new{\DG}  {{\ifmmode \Delta\Gamma_s       \else $\Delta\Gamma_s$ \fi}}

\begin{document}

\title{\Bs Mixing and Lifetime Difference at CDF } 

%

\author{Sinead M. Farrington (for the CDF collaboration)}
\affiliation{University of Oxford, Oxford, OX1 3RH, UK}
\begin{abstract}
The measurements of the \Bs mixing parameter which describes the difference in width (the inverse of the lifetime)
between the \Bs heavy and light states are described.  These measurements include direct measurement
of $\Delta \Gamma_s$ and $\Gamma_s$ by resolving CP eigenstates by angular analysis
and measurements in which the proportion of CP even and odd is known.
\end{abstract}

\maketitle

\thispagestyle{fancy}


\section{THE \Bs MESON SYSTEM} 

The \Bs meson is a bound state consisting of an anti-bottom quark and a strange quark. 
Its antimatter partner consists of a bottom quark and an anti-strange quark.  These 
two mesons can be interchanged by a second order weak interaction, represented as a
Feynman box diagram.  The parameters which describe this mixing between \Bs and \Bsbar
are predictable parameters in the  Standard Model which could be modified by non 
Standard Model contributions to the box diagram.  The system is usually characterised
in terms of heavy and light mass eigenstates which are superpositions of the \Bs and 
\Bsbar and thus there are  five key parameters:  masses ($m_H$ and $m_L$), widths 
($\Gamma_H$ and $\Gamma_L$) and a phase ($\beta_s$).  The first measurement of the difference
in mass, $\Delta m_s$, was made in 2006 \cite{mixmeas}.  Since this measurement is so precise, the B physics program
at CDF now focuses on the measurement of $\Delta \Gamma = \Gamma_L-\Gamma_H$ which is 
discussed here, $\beta_s$  and $\Gamma=\Gamma_H+\Gamma_L /2$ 
discussed in the proceedings of D. Tonelli and S. Behari from this conference.

\section{MEASUREMENT OF $\Delta\Gamma$}
While $\Delta m_s$ has been measured to great precision, \DG has so far been measured imprecisely.  
A measurement of \DG provides an extra test of the Standard Model since new physics may enter
through the phase $\beta_s$:
\begin{equation}
\Delta\Gamma=\Delta\Gamma^{SM}\times |\cos 2\beta_s|
\end{equation}
With measurements of both $\Delta m_s$ and $\Delta\Gamma$, the following Standard Model relation can be tested, 
where $B,\space B_s$ and $B_R$ are parameters from lattice QCD \cite{dgrel}:
\begin{equation}
\frac{\Delta\Gamma_s}{\Delta m_s}=\left[\left(46.2\pm4.4\right)+\left(10.6\pm1.0\right)\times \frac{B_s}{B}-\left(11.9\pm1.3\right)\times \frac{B_R}{B}\right]\times 10^{-4}
\end{equation}
In interpreting the results presented in this report, the assumption should be made 
that the \Bs light mass eigenstate is CP even and the heavy state is CP odd.  Two approaches
to measuring \DG are pursued.  The first is to analyse $\Bs\to J/\psi \phi$ decays, fitting
the angular distributions between the decay products in order to decipher the CP odd and even content.  The second is to measure the width in a CP specific decay for which the proportion
of CP odd and even states is known a priori.

\subsection{Direct Measurement of \DG: $\Bs \to J/\psi \phi$}
\begin{figure}[htb]
\includegraphics[scale=0.4]{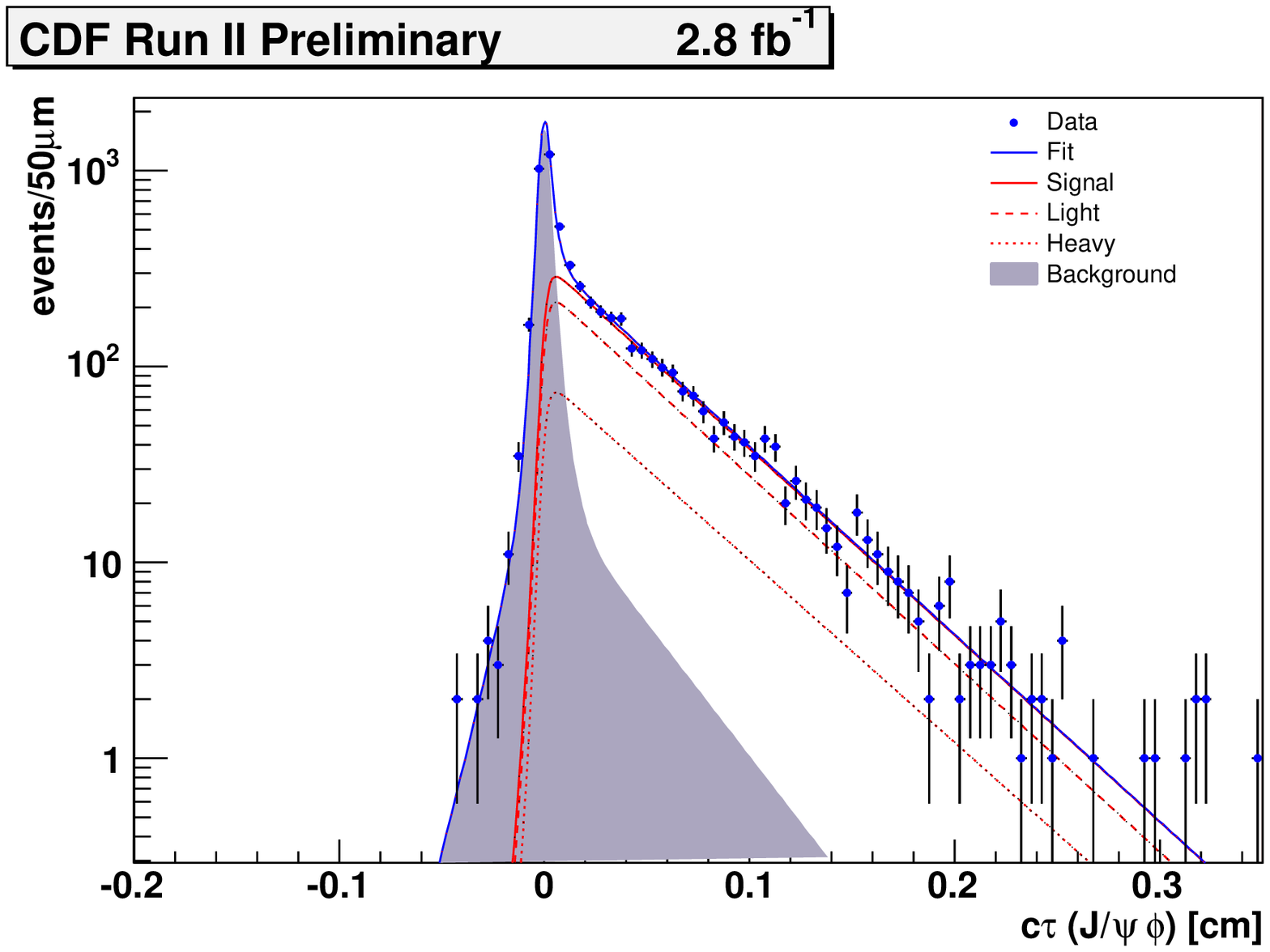}
\caption{\it Lifetime distribution of $\Bs\to J/\psi\phi$ candidates with fit superimposed.}
\label{fig:jpsiphi}
\end{figure}

Approximately 3150 decays of $\Bs\to J/\psi \phi$ are gathered in 2.8 fb$^{-1}$ of CDF Run II data.  In this sample,
a simultaneous fit is made to the mass, lifetime and angular variables.  The lifetime distribution and its fit projection
are shown in Figure \ref{fig:jpsiphi}.  Since the decays are 
of a pseudo scalar decaying into two vector mesons, the decays contain S and D wave states which
are CP even and P wave states which are CP odd.  Therefore angular distributions in the
$J/\psi$ and $\phi$ rest frames give information on the CP composition of the $\Bs\to J/\psi\phi$ decays, measured
in terms of $|A_0|$ and $|A_\parallel|$ which indicate the S,D and P wave composition.
The results of the simultaneous fit assuming no CP violation are 
\begin{eqnarray}
c \tau_s&=&459\pm 12(stat) \pm 3(sys)\hspace{1mm}\mu m\nonumber\\
\DG&=&0.02\pm0.05(stat)\pm 0.01(sys) \hspace{1mm} ps^{-1}\nonumber\\
|A_0|^2&=&0.508\pm0.024(stat)\pm 0.008(sys)\nonumber\\
|A_\parallel|^2&=& 0.241\pm0.019(stat)\pm0.007(sys)\nonumber
\end{eqnarray}
The main systematic uncertainties in the \DG measurement are from the lifetime resolution model and 
the background lifetime model.  The result can be compared to the predicted value of $\DG=0.096\pm 0.039\hspace{1mm} ps^{-1}$ \cite{dgpred}.  The previous iteration of this analysis has been published \cite{dgmeas}.

\subsection{Measurement of \DG from CP specific state: $\Bs\to D^+_s D^-_s$}
\begin{figure}[htb]
\includegraphics[scale=0.5]{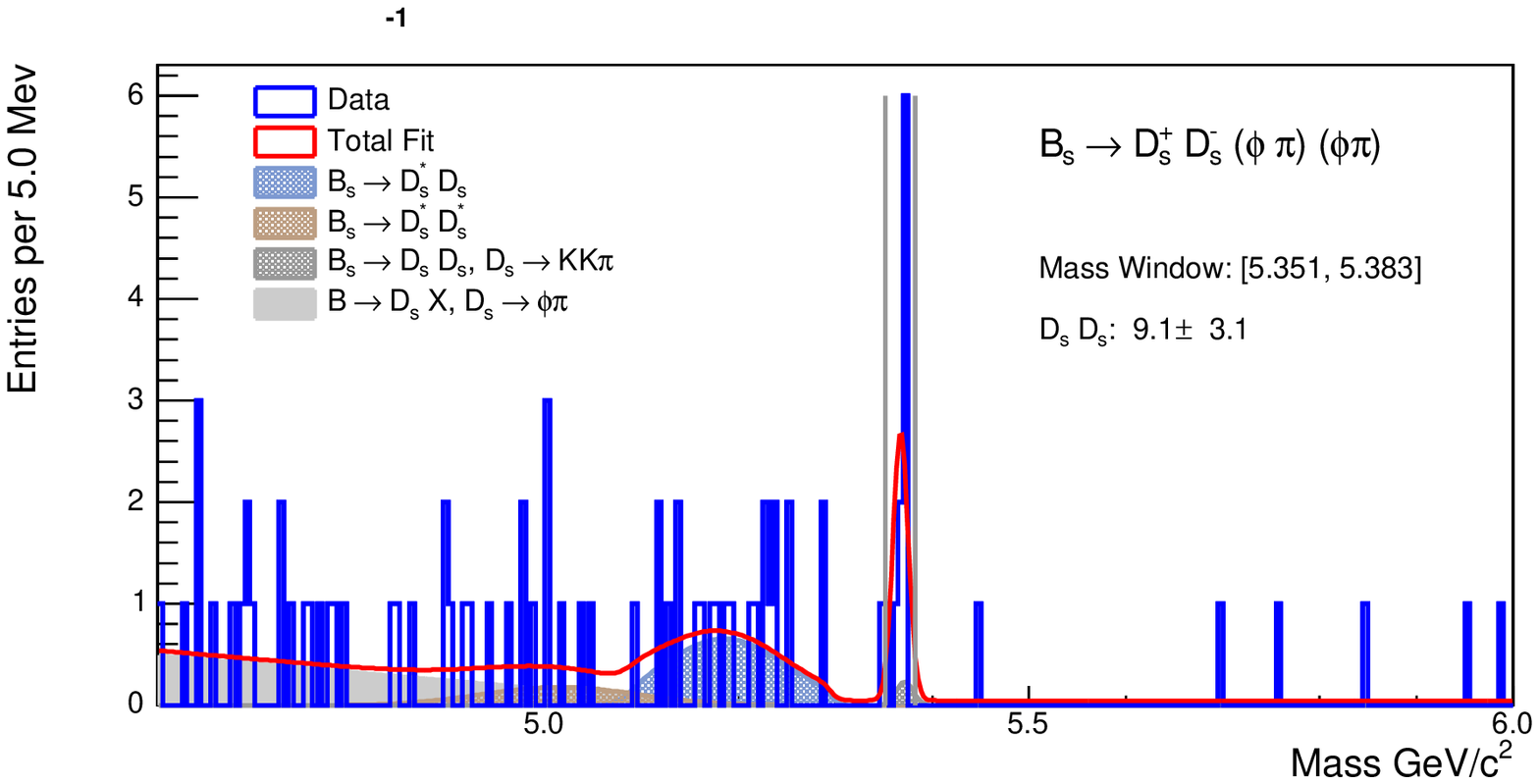}
\caption{\it $\Bs \to D_s^+ D_s^-$ candidates with mass fit superimposed.}
\label{fig:bdsds}
\end{figure}

The decay $\Bs\to D^+_s D^-_s$ is a $b\to c\overline{c}s$ decay which is purely CP even 
in composition.  Therefore a width measurement of $\Bs\to D^+_s D^-_s$ would measure
$\Gamma_L$.  The branching ratio of the $\Bs\to D_s^+ D_s^-$ mode is related to the difference in width between the two weak eigenstates \cite{dgamma} and so can provide an indirect measurement of $\Delta \Gamma_s$:
\begin{equation}
\frac{\DG}{\Gamma_s}=2\times BR\left(\Bs\to D^{(*)+}_s D^{(*)-}_s\right)
\label{eqn:dsds}
\end{equation}
The assumption is made that the width difference is mainly caused by $\Bs\to D^{(*)+}_s D^{(*)-}_s$ decays.  Other decays
which are not purely CP even are assumed to have much smaller branching fractions.  The $\Bs \to D_s^+ D_s^-$ ($D_s^+\to\phi\pi$,$D_s^-\to \phi\pi$ or $K^*K$ or $\pi\pi\pi$) branching ratio is measured relative to that of $B^0_d\to D_s^+ D^-$ in order to eliminate sources of systematic uncertainty.  The remaining candidates with $D_s^+\to\phi\pi^+$ after selection are shown in Figure \ref{fig:bdsds} for the $\Bs \to D_s^+ D_s^-$ mode, which is clean owing to the presence of two narrow resonance $\phi$ mesons in the final state of the $D^+_s$ decay.  The mass distributions are fitted with a three component fit: the Gaussian signal distribution; the combinatorial background distribution obtained by fitting the high sideband and extrapolating it under the peak; and the physics backgrounds, obtained from Monte Carlo simulations.  A multi-parameter fit is performed, yielding greater than five standard deviations significance and a branching ratio of \cite{bsdsds}:
\begin{eqnarray}
\frac{BR(\Bs \to D_s^+ D_s^-)}{BR(B^0_d\to D_s^+ D^-)} &=& 1.44^{+0.38}_{-0.31} (stat)^{+0.08}_{-0.12}(syst)
                 \pm 0.21\left(\frac{f_s}{f_d}\right)\pm0.20(BR(\phi\pi)) \nonumber
\end{eqnarray}
From this measurement, a 95\% confidence level limit of $\DG/\Gamma_s\geq 0.012$ can be set using equation \ref{eqn:dsds}.  A lower limit is set since equation \ref{eqn:dsds} includes excited $D_s$ states while this measurement is made on the 
non-excited $D_s$ states.

\subsection{Measurement of \DG from CP specific state: $\Bs \to K^+ K^-$}
\begin{figure}[htb]
\includegraphics[scale=0.3]{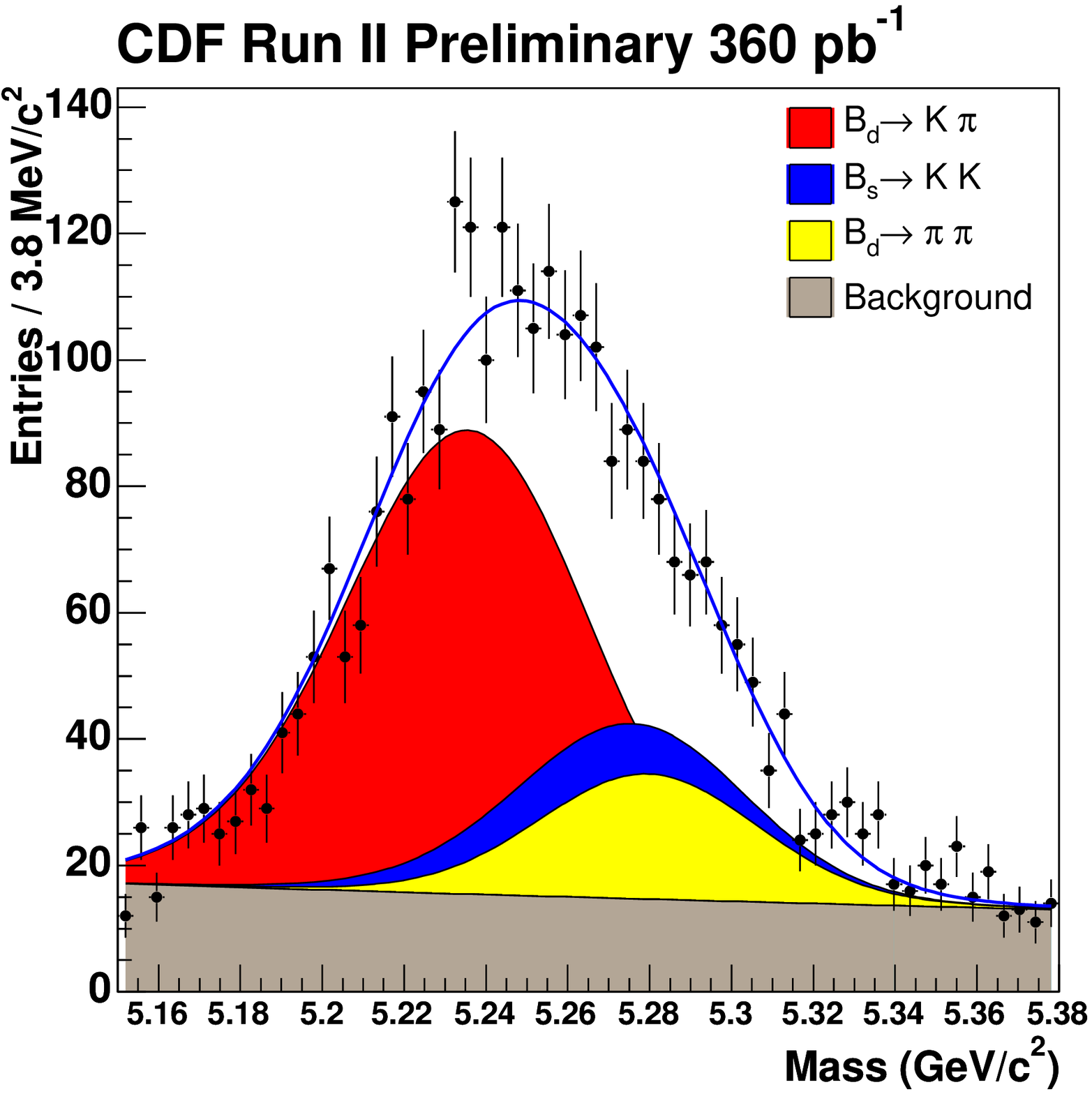}
\caption{\it $\Bs \to K^+ K^-$ candidates with mass fit superimposed.}
\label{fig:kkmass}
\end{figure}

Fully hadronic decays of B mesons can be resolved at CDF thanks to its displaced track trigger which requires a pair of tracks 
in the 
drift chamber at level 1 in the triggering system, and a displaced track in the silicon detector at level 2.
In addition, the CDF detector has excellent mass resolution which aids in separation of the various B
decays into pairs of hadrons.  The projection of the fit to invariant mass, particle identification quantities and 
a measure of momentum imbalance is shown in Figure \ref{fig:kkmass}.
The decay $\Bs \to K^+ K^-$ is almost 100\% CP even and therefore a measurement of the \Bs meson lifetime in
this state can be combined with knowledge of the average \Bs lifetime in order to obtain a measurement of
\DG.  A lifetime fit is performed in 0.36 fb$^{-1}$ of CDF Run II data and yields
\begin{equation}
\tau(\Bs\to K^+ K^-)=1.53\pm0.18(stat)\pm 0.02(sys)\hspace{1mm} ps
\end{equation}
Using the Heavy Flavour Averaging Group flavour specific lifetime, $\tau(\Bs)=1.454\pm0.040$ ps, yields
\begin{equation}
\DG/\Gamma_s=-0.08\pm0.23\pm0.03 
\end{equation}
\section{CONCLUSIONS}
Several measurements of \DG have been presented.  In the presence of new physics these measurements
need not necessarily agree.  With increased precision in Tevatron analyses, hints of new physics may be revealed.


\begin{thebibliography}{9}   
\bibitem{mixmeas}
A.Abulencia et al., The CDF Collaboration, Phys. Rev. Lett. 97, 242003 (2006)
\bibitem{dgrel}
A. Lenz, U. Nierste, JHEP 06, 072 (2007)
\bibitem{dgpred}
A. Lenz, arXiv 0802.0977 (2008)
\bibitem{dgmeas}
T. Aaltonen et al., The CDF Collaboration, Phys. Rev. Lett. 100, 121803 (2008)
\bibitem{dgamma} I. Dunietz et al., hep-ph/0012219
\bibitem{bsdsds} T. Aaltonen et al., The CDF Collaboration, Phys. Rev. Lett. 100, 021803 (2008)
\end{thebibliography}
\end{document}